\documentclass[aps,twocolumn,showpacs,preprintnumbers,amsmath,amssymb,superscriptaddress,floatfix,nofootinbib]{revtex4}

\usepackage{graphicx}
\usepackage{amsmath}
\usepackage{amsfonts}
\usepackage{amssymb}
\usepackage{color}

\newcommand{\Slash}[1]{\ooalign{\hfil/\hfil\crcr$#1$}}

\begin{document}

\title{Effects of a $N^*_{c\bar{c}}$ resonance with hidden charm in the $\pi^- p \to D^- \Sigma_c^+$ reaction near threshold}

\author{E. J. Garzon}
\affiliation{Institute of Modern Physics, Chinese Academy of
Sciences, Lanzhou 730000, China} \affiliation{Departamento de
F\'{\i}sica Te\'orica and IFIC, Centro Mixto Universidad de
Valencia-CSIC, Institutos de Investigaci\'on de Paterna, Aptdo.
22085, 46071 Valencia, Spain}

\author{Ju-Jun Xie}~\email{xiejujun@impcas.ac.cn}
\affiliation{Institute of Modern Physics, Chinese Academy of
Sciences, Lanzhou 730000, China} \affiliation{Research Center for
Hadron and CSR Physics, Institute of Modern Physics of CAS and
Lanzhou University, Lanzhou 730000, China} \affiliation{State Key
Laboratory of Theoretical Physics, Institute of Theoretical Physics,
Chinese Academy of Sciences, Beijing 100190, China}

\begin{abstract}

We study the effect of a hidden charm nuclear excited state
$N^*_{c\bar{c}}$ in the $\pi^- p \to D^- \Sigma_c^+$ reaction near
threshold using an effective Lagrangian approach. We calculate the
background contribution of the $t$ and $u$ channels by the $D^{*0}$
vector meson exchange and $\Sigma_c^{++}$ intermediate state,
respectively. We show that the consideration of a $N^*_{c\bar{c}}$
resonance provides an enhancement of the total cross section close
to the reaction threshold. We also evaluate the differential cross
section for different energies and we study the angle dependence. It
is expected that our model calculations will be tested in future
experiments.

\end{abstract}

\date{\today}


\maketitle

\section{Introduction}

The study of the hidden charm sector has been a very active field
lately with many studies involving interactions of meson-baryon or
meson-meson~\cite{Lutz:2005vx,Liang:2014kra,Liang:2013laa,Albaladejo:2013aka,Garcia-Recio:2013gaa,Uchino:2015uha,Garcia-Recio:2015jsa}.
In particular the $N^*_{c\bar{c}}(4261)$ resonance is predicted in
many models including the meson-baryon
interaction~\cite{Wu:2010vk,Wu:2010jy} or heavy quark spin
symmetries~\cite{Xiao:2013yca}. The role of those states in the
production of charmed hadrons is a new field that requires an
exhaustive study because new facilities will have enough energy to
generate those particles. One example is the pion beam experiments
at J-PARC where the energy of the pion is expected to be over 20
GeV~\cite{Kim:2014qha}, and therefore, it is sufficient to produce
those hidden charmed baryons at J-PARC. In the near future the
J-PARC in Japan will be one of the efficient facilities in which to
study the predicted baryon states dynamically generated from the
meson-baryon interaction with hidden charm.

The effective Lagrangian approach has been successfully used in
Ref.~\cite{Wu:2014yca} where three $N^*$ resonances are included to
reproduce the $\pi^- p \to K^0 \Lambda$ reaction near the threshold
and also in Ref.~\cite{Xie:2015zga} which is a study about the role
of the $\Lambda_c^+$(2940) state in the $\pi^- p \to D^- D^0 p$. In
Refs.~\cite{Wu:2010vk,Wu:2010jy,Xiao:2013yca}, it was found that the
$N^*_{c\bar{c}}(4261)$ couples mostly to $\bar{D}\Sigma_c$, and it
could be considered a $\bar{D}\Sigma_c$ bound state that, however,
decays into the opened channels $\eta_c N$ and $J/\psi N$. In
Ref.~\cite{Wu:2010jy}, the production cross sections of the
$N^*_{c\bar{c}}(4261)$ resonance in the $\bar{p}p \to \bar{p}p
\eta_c$ and $\bar{p}p \to \bar{p}p J/\Psi$ reactions were estimated.
In Ref.~\cite{Wang:2015qia}, the $N^*_{c\bar{c}}(4261)$ resonance is
considered in the $\pi^- p \to \eta_c n$ reaction, where a clear
signal is expected in the total cross section of the $\pi^- p \to
\eta_c n$ reaction. Huang \textit{et al.}~\cite{Huang:2013mua},
studied the photo-production of this $N^*_{c\bar{c}}(4261)$
resonance and they found that it is difficult to search for the
$N^*_{c\bar{c}}(4261)$ resonance in photo-production because of the
very large background.

It has been found in previous
works~\cite{Gamermann:2007mu,MartinezTorres:2012du,Xie:2013ula} that
states below the threshold of a process can generate bumps close to
the threshold, leading to claims of resonance above the threshold of
the reaction studied; hence, we expect that if the
$N^*_{c\bar{c}}(4261)$ state is created in the $\pi^- p$ scattering,
a strong signal in the $D^- \Sigma_c^+$ production should be visible
close to the reaction threshold.

In this work, we study the role of the hidden charm nuclear excited
state $N^*_{c\bar{c}}(4261)$ ($\equiv N^*_{c\bar{c}}$) in the $\pi^-
p \to D^- \Sigma^+_c$ reaction close to the threshold using the
effective Lagrangian approach. In addition to the contribution from
the $s$ channel hidden charm $N^*_{c\bar{c}}$ resonance, the
background contributions from the $t$ channel $D^{*0}$ exchange and
the $u$ channel $\Sigma^{++}_c$ exchange are also considered. We
calculate the total and differential cross sections of the $\pi^- p
\to D^- \Sigma^+_c$ reaction near the threshold.

This article is organized as follows. In Sec. II we present the
formalism and the procedure in our calculations. We give the results
and some discussion in Sec. III. Finally in Sec. IV we give some
conclusions.

\section{Formalism}

\begin{figure*}[htbp]
\begin{center}
\includegraphics[scale=0.6]{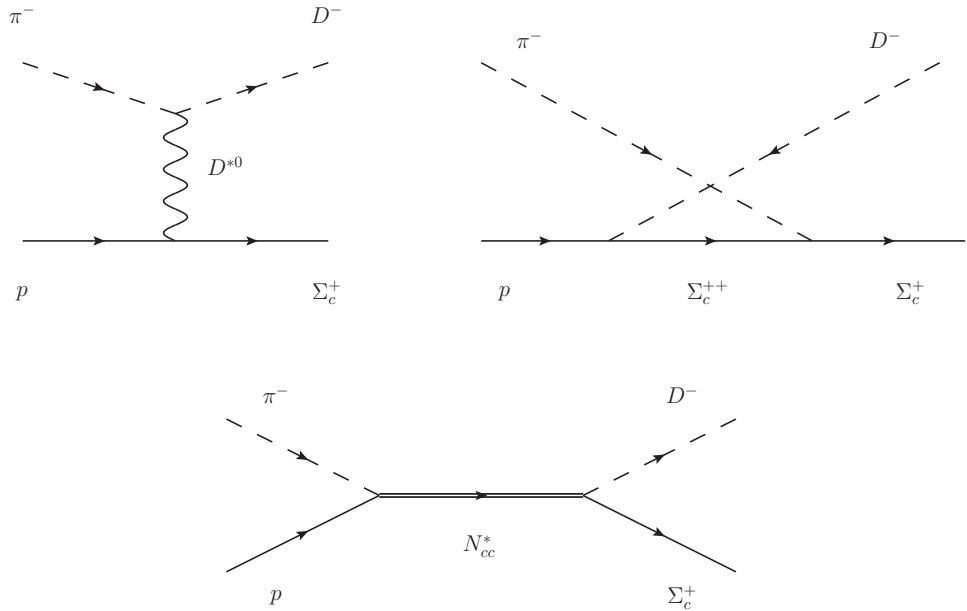}
\caption{Diagrams considered in the $\pi^- p \to D^- \Sigma_c^+$
reaction.} \label{fig:diagrams}
\end{center}
\end{figure*}

We study the $\pi^- p \to D^- \Sigma_c^+$ reaction using an
effective Lagrangian approach as in Ref.~\cite{Xie:2015zga} and in
many other
works~\cite{Dong:2010xv,Tsushima:1998jz,Sibirtsev:2005mv,Liu:2006tf,Maxwell:2012zz,
Liu:2012kh,Xie:2013wfa,Liu:2011sw,Liu:2012ge,Lu:2013jva,Lu:2014rla,Xie:2007qt,Xie:2013mua,Xie:2010yk}.
We want to study the effect of the hidden charm
$N^*_{c\bar{c}}(4261)$ resonance in the cross section of the $\pi^-
p \to D^- \Sigma_c^+$ reaction. To see the effect of this resonance
has on the total cross section, we consider the background of the
reaction. As shown in Fig.~\ref{fig:diagrams} we are going to
include (a) the $t$-channel mediated by a $D^{*0}$ vector meson
exchange and (b) the $u$-channel considering a $\Sigma_c^{++}$ state
as an intermediate state. We include the $s$-channel assuming the
creation of a $N^*_{c\bar{c}}$ state in the process, as shown in
Fig.~\ref{fig:diagrams}(c).

To evaluate the diagrams shown in Fig.~\ref{fig:diagrams}, we use
the effective Lagrangian densities of the interaction vertices. We
use the Lagrangian densities of $D^*D\pi$ $\Sigma_c p D^{*}$,
$N^*_{c\bar{c}} N \pi$, $N^*_{c\bar{c}} \Sigma_c \bar{D}$, $D N
\Sigma_c$, and $\pi \Sigma_c \Sigma_c$ as follows,
\begin{eqnarray}
\mathcal{L}_{D^* D \pi} &=& g_{D^*D\pi}  \bar{D}^{*\mu}  \left(D \partial_\mu \pi - \pi \partial_\mu D \right), \\
\mathcal{L}_{\Sigma_c p D^{*}} &=& g_{\Sigma_c p D^{*}} \bar{\Sigma}_c \gamma_\mu N D^{*\mu} + {\rm h.c.}, \\
\mathcal{L}_{N^*_{c\bar{c}} N \pi}&=& -g_{N^* N \pi} \bar{N}^* \pi N  + {\rm h.c.},\\
\mathcal{L}_{N^*_{c\bar{c}} \Sigma_c \bar{D}}&=&-g_{N^* \Sigma_c D} \bar{N}^* D \Sigma_c + {\rm h.c.},\\
\mathcal{L}_{D N \Sigma_c}&=&-ig_{D N \Sigma_c} \bar{N} \gamma_5 D \Sigma_c + {\rm h.c.},\\
\mathcal{L}_{\pi \Sigma_c \Sigma_c}&=&-ig_{\pi \Sigma_c \Sigma_c}
\bar{\Sigma}_c \gamma_5 \pi \Sigma_c + {\rm h.c.}.
\end{eqnarray}

From Ref.~\cite{Wu:2010jy} we have a prediction for the partial
decay width of the $N^*_{c\bar{c}}(4261)$ resonance to the $\pi N$
channel, $\Gamma \left(N^*_{c\bar{c}} \to \pi N \right) = 3.8$ MeV.
We can get the coupling constant, $g_{N^* N \pi}$, from the formula
of the partial decay width of the $N^*_{c\bar{c}}(4261)$ resonance
to this channel,
\begin{eqnarray}
\Gamma \left( N^* \to \pi N \right) &=& \frac{3 g^2_{\pi N
N^*}}{4\pi} \frac{E_N+m_N}{M_{N^*}}|\vec{p}_N|, \\
E_N &=& \frac{M^2_{N^*} + m^2_N - m^2_{\pi}}{2M_{N^*}}, \\
|\vec{p}_N| &=& \sqrt{E^2_N - m^2_N},
\end{eqnarray}
with $\vec{p}_N$ being the three-momentum of final proton in the
rest frame of the $N^*_{c\bar{c}}(4261)$ resonance. Using the value
$M_{N^*} = 4261.87$ MeV for the mass of the $N^*_{c\bar{c}}(4261)$
resonance, we get a coupling of $g_{\pi N N^*} = 0.1$. In
Ref.~\cite{Xiao:2013yca} there is a calculation for the coupling of
the $N^*_{c\bar{c}}(4261)$ resonance to the $\bar{D} \Sigma_c$ being
$g_{N^* \Sigma_c \bar{D}} = 3.13$. In the case of $g_{D^*D\pi}$, as
done in Ref.~\cite{Xie:2015zga}, we assume the same partial decay
width as for $D^{*0}\to D^0 \pi^0$, with a coupling of $g_{D^*D\pi}
= 14.1$.

Furthermore, the coupling constants $g_{\Sigma_c N D^*} = -7.8$,
$g_{D N \Sigma_c} = 2.69$, and $g_{\pi \Sigma_c \Sigma_c} = 10.76$
are determined from $SU(4)$ invariant
Lagrangians~\cite{Dong:2010xv,Liu:2001ce,Okubo:1975sc} and $SU(3)$
flavor symmetry~\cite{Doring:2010ap,de Swart:1963gc} in terms of
$g_{\pi NN} = 13.45$ and $g_{\rho NN} = 6$. We list in
Table~\ref{tabmass} also the mass and spin-parity of the other
related states in our calculation.

\begin{table}[htbp]
\begin{center}
\caption{\label{tabmass} Parameters used in the present
calculation.}
\begin{tabular}{ccc}
\hline
State       & Mass (MeV)  &  Spin-parity ($J^P$)    \\
\hline
$\pi^-$       & $139.57$   & $0^-$   \\
$p$    & $938.27$   & $\frac{1}{2}^+$  \\
$D^-$ & $1869.61$ & $0^-$    \\
$\Sigma^{+}_c$    & $2452.90$   & $\frac{1}{2}^+$  \\
$D^{*0}$ & $2006.96$ & $1^-$    \\
\hline
\end{tabular}
\end{center}
\end{table}

The evaluation of the diagrams of Fig.~\ref{fig:diagrams} using the
Lagrangian densities shown above, leads us to the following
amplitudes for the $t$, $u$ and $s$ channels,
\begin{eqnarray}
\mathcal{M}_t  &=& - \frac{g_{D^* D \pi} g_{\Sigma_c N D^{*}}}{q_t^2
- M_{D^{*0}}^2} \times \nonumber \\
&&  \bar{u}_{\Sigma_c^+}(p_f) (\Slash
p_{\pi} - \frac{q_{t} \cdot p_{\pi} \Slash q_{t}}{M_{D^{*0}}^2}) u_p(p_i),   \\
\mathcal{M}_u &=& - \frac{g_{D N \Sigma_c} g_{\pi \Sigma_c
\Sigma_c}}{q^2_u - m_{\Sigma_c^{++}}^2} \times \nonumber \\
&& \bar{u}_{\Sigma_c^+}(p_f) (\Slash q_u + m_{\Sigma_c^{++}}) u_p(p_i), \\
\mathcal{M}_s &=& \frac{\sqrt{2} g_{N^* N \pi} g_{N^* \Sigma_c
\bar{D}}}{q^2_s - M^2_{N^*}} \times \nonumber \\
&& \bar{u}_{\Sigma_c^+}(p_f) (\Slash q_s - M_{N^*}) u_p(p_i),
\end{eqnarray}
where $p_i$ and $p_f$ are four-momenta for the initial proton and
final $\Sigma_c^+$, respectively. In the above equations,
$\mathcal{M}_t$ corresponds to Fig.~\ref{fig:diagrams}(a), and
$\mathcal{M}_u$ and $\mathcal{M}_s$ correspond to
Figs.~\ref{fig:diagrams} (b) and \ref{fig:diagrams} (c),
respectively. In the equations above $q_i (i = s,t,u)$ represents
the four momentum of the particle exchanged in each of the channels.
In the $t$-channel $q_t$ is the four-momentum of the $D^{*0}$ which
corresponds to $q_t^2=(p_\pi - p_D)^2$ equivalent to the Mandelstam
variable $t$, for the $u$-channel $q_u$ is the four-momentum of the
$\Sigma_c^{++}$ being $q^2_u = (p_\pi - p_{\Sigma_c})^2$ or $u$, and
for the $s$-channel $q_s$ is the four-momentum of the
$N^*_{c\bar{c}}(4261)$, and $q^2_s = (p_\pi + p_p)^2 = s$ is the
invariant mass square of the $\pi^- p$ system.

Besides, we need to add form factors for the hadrons since they are
not point-like particles. In the case of the $D^{*0}$ meson, we use
the form factor used in
Refs.~\cite{He:2011jp,Dong:2014ksa,Oh:2007jd,Haidenbauer:2009ad} as
follows
\begin{equation}
F_{D^*}(q^2_t,M_{D^*})=\frac{\Lambda^2_{D^*} -
M^2_{D^*}}{\Lambda^2_{D^*}-q^2_{t}}
\end{equation}
In the case of the baryons, we use another form factor as done in
Refs.~\cite{Feuster:1997pq,Shklyar:2005xg}
\begin{equation}
F_B(q^2_{\rm{ex}},M)=\frac{\Lambda_B^4}{\Lambda^4_B + (q^2_{\rm{ex}}
- M^2)^2}
\end{equation}
where $M$ is the mass of the exchanged baryon $m_{\Sigma_c}$ and
$M_{N^*_{c\bar{c}}}$, and $q_{\rm{ex}}$ is the exchanged
four-momentum of each baryon. In our study we use all the cut off
parameters $\Lambda = \Lambda_{D^*} = \Lambda_{\Sigma_c} =
\Lambda_{N^*_{c\bar{c}}} = 2.5$ GeV to minimize the free parameters.

The differential cross section in the center of mass (${\rm c.m.}$)
frame for the $\pi^- p \to D^- \Sigma_c^+$ reaction is calculated
using the following equation
\begin{equation}
\frac{d \sigma}{d\Omega}=\frac{d \sigma}{2\pi d \text{cos} \theta}=
\frac{m_{\Sigma_c} m_p}{32 \pi^2 s}
\frac{|\vec{p}\,^{\rm{c.m.}}_D|}{|\vec{p}\,^{\rm{c.m.}}_\pi|}
{\sum}|\mathcal{M}_{\pi^- p \to D^- \Sigma_c^+}|^2,
\end{equation}
where $\mathcal{M}_{\pi^- p \to D^- \Sigma_c^+} = {\cal M}_t + {\cal
M}_u + {\cal M}_s$ is the total scattering amplitude of the $\pi^- p
\to D^- \Sigma_c^+$ reaction, and $\theta$ is the scattering angle
of the outgoing $D^-$ meson relative to the beam direction, while
$\vec{p}\,^{\rm{c.m.}}_\pi$ and $\vec{p}\,^{\rm{c.m.}}_D$ are the
$\pi^-$ and $D^-$ three momenta in the c.m. frame, which are
\begin{eqnarray}
|\vec{p}\,^{\rm{c.m.}}_{\pi}| &=&
\frac{\lambda^{1/2}(s,m^2_{\pi},m^2_p)}{2\sqrt{s}}, \\
|\vec{p}\,^{\rm{c.m.}}_D| &=&
\frac{\lambda^{1/2}(s,m^2_{D},m^2_{\Sigma_c})}{2\sqrt{s}},
\end{eqnarray}
where $\lambda$ is the K\"allen function with $\lambda(x,y,z) =
(x-y-z)^2 - 4yz$.

\section{Numerical Results}

\begin{figure}[htbp]
\begin{center}
\includegraphics[scale=0.7]{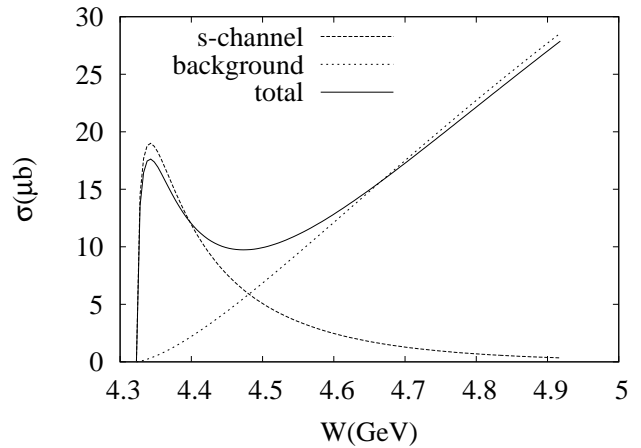}
\end{center}
\caption{Comparison of the total cross section of the $\pi^- p \to
D^- \Sigma_c^+$ reaction: (solid line) total cross section including
$N^*_{c\bar{c}}(4261)$, (dotted line) cross section of the
background ($t$ and $u$ channels) and (dashed line) cross section
for the $s$ channel [$N^*_{c\bar{c}}(4261)$] only.}
\label{fig:cross}
\end{figure}

In this section, we show our theoretical results for the total and
differential cross section of the $\pi^- p \to D^- \Sigma_c^+$
reaction near the threshold. We have evaluated the diagrams of the
$\pi^- p \to D^- \Sigma_c^+$ reaction with a $D^{*0}$ exchange in
the $t$ channel and with $\Sigma_c^{++}$ as an intermediate state in
the $u$ channel. Those diagrams provide us the background of the
reaction where we can see the effects of the $N^*_{c\bar{c}}(4261)$
resonance under the threshold, then we are able to compare the
effect in the cross section. In Fig.~\ref{fig:cross} we show our
numerical results for the total cross section as a function of the
invariant mass $W = \sqrt{s}$ of the $\pi^- p$ system comparing the
effect of including or not the $N^*_{c\bar{c}}(4261)$ resonance in
the total scattering amplitude.

As we can see in Fig.~\ref{fig:cross}, the dotted line shows the
cross section of the background including the $t$ channel mediated
by the exchange of a $D^{*0}$ meson and the $u$ channel where
$\Sigma_c^{++}$ is considered an intermediate state. The dashed line
includes only the contribution of the $s$ channel process with the
$N^*_{c\bar{c}}(4261)$ resonance; we can see the important
contribution of this state to the total cross section of the $\pi^-
p \to D^- \Sigma_c^+$ reaction close to the reaction threshold where
there is a clear enhancement.

\begin{figure}[htbp]
\begin{center}
\includegraphics[scale=0.7]{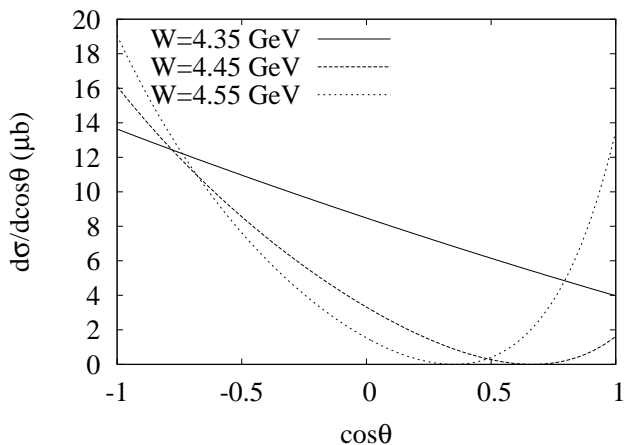}
\end{center}
\caption{Differential cross section of the $\pi^- p \to D^-
\Sigma_c^+$ reaction at different energies in center of mass frame:
$W = 4.35$ GeV (solid line), $W = 4.45$ GeV (dashed line) and $W =
4.55$ GeV (dotted line).} \label{fig:dcs}
\end{figure}

In addition to the total cross section, we present in
Fig.~\ref{fig:dcs} the differential cross section of this reaction
depending on the scattering angle $\theta$ for different energies.
We show the results for $W = 4.35$ GeV (solid line) because this
energy point is close to the peak in the total cross section and
should be dominated by the $s$ channel $N^*_{c\bar{c}}(4261)$
resonance. This is what occurs but we can see in
Fig.~\ref{fig:cross} a small contribution of the background at this
energy point. In the case of other energy points, with $W = 4.45$
GeV (dashed line) we have a mix of the background and the $s$
channel $N^*_{c\bar{c}}(4261)$ resonance and we can see this effect
in Fig.~\ref{fig:dcs} where the dependence on the angle starts to be
relevant. Finally, with $W = 4.55$ GeV (dotted line), the background
almost dominates the behavior and the dependence on the angle of the
differential cross section becomes more important. This phenomenon
with the clear threshold enhancement of the total cross section
shown in Fig.~\ref{fig:cross} shows that the contributions of the
$N^*_{c\bar{c}}(4261)$ resonance and the background are sizably
different. We hope that this feature may be used to study the
$N^*_{c\bar{c}}(4261)$ resonance in future experiments.

\section{Conclusions}

We have studied the total and differential cross sections of the
$\pi^- p \to D^- \Sigma_c^+$ reaction near the threshold and the
effects of the presence of a $N^*_{c\bar{c}}$ resonance. The
background of this reaction is taken into account by the exchange of
a $D^{*0}$ meson and $\Sigma_c^{++}$ as an intermediate state while
we consider the $s$ channel mediated by the $N^*_{c\bar{c}}(4261)$
resonance. We use an effective Lagrangian approach to calculate the
interaction vertices for the considered diagrams and where the
different couplings of the Lagrangians are determined using partial
decay widths or $SU(4)$ Lagrangian relations. The evaluation of the
$t$ and $u$ channels provides us a background to study the effects
of the consideration of the $N^*_{c\bar{c}}(4261)$ state. The
results of the total cross section show a clear enhancement close to
the reaction threshold that could be observed in future experiments
with pion beams. We also evaluate the differential cross sections
for different energies and we predict s small dependence on the
angle close to the threshold but the dependence starts to be
important as the energy increases and the background contribution
becomes dominant. We hope that this study helps us to understand the
results of future experiments at J-PARC.

\section*{Acknowledgments}

This work is partly supported by the National Natural Science
Foundation of China under Grant No. 11475227.

\bibliographystyle{unsrt}

\end{document}